\documentclass{article}

    \PassOptionsToPackage{numbers, compress}{natbib}

\usepackage[final]{neurips_2019}

\usepackage[]{neurips_2019}

\usepackage[utf8]{inputenc} 
\usepackage[T1]{fontenc}    
\usepackage{hyperref}       
\usepackage{url}            
\usepackage{booktabs}       
\usepackage{amsfonts}       
\usepackage{nicefrac}       
\usepackage{microtype}      
\usepackage{graphicx}
\usepackage{siunitx}

\usepackage{amsmath}
\newcommand\raisepunct[1]{\,\mathpunct{\raisebox{0.5ex}{#1}}}
\usepackage{graphicx}
\usepackage[font=scriptsize]{caption}
\usepackage{subcaption}

\title{Measuring Impact of Climate Change on Tree Species: analysis of JSDM on FIA data}

\author{%
    Hyun Choi, 
    Ali Sadeghian,
    Sergio Marconi,
    Ethan White,
    Daisy Zhe Wang
    \\
    University of Florida \\
    Gainesville, FL \\
  \texttt{\{hchyun, asadeghian, 
daisyw\}@ufl.edu}, \texttt{\{ethan, 
sergio.marconi\}@weecology.org} \\
}

\begin{document}

\maketitle

\begin{abstract}
One of the first beings affected by changes in the climate are trees, one of our most vital resources. In this study tree species interaction and the response to climate in different ecological environments is observed by applying a joint species distribution model to different ecological domains in the United States. Joint species distribution models are useful to learn inter-species relationships and species response to the environment. The climates’ impact on the tree species is measured through species abundance in an area. We compare the model’s performance across all ecological domains and study the sensitivity of the climate variables. With the prediction of abundances, tree species populations can be predicted in the future and measure the impact of climate change on tree populations.
\end{abstract}

\section{Introduction}

Ecologist expect the rapid change in climate to have unprecedented impacts on species geographical redistribution. Shifts in economically important species will likely have significant impacts on human well-being, forest management strategies and conservation planning \cite{araujo2007importance,hsiang2017estimating,rolnick2019tackling}.
Reliably predicting future patterns in species distributions has become a major goal in ecology, and hence numerous species distribution models have been developed \cite{araujo2007importance,iverson2019analysis,boisvert2019divergent}. Modern implementations use historical climate data and a few topographic variables to predict individual species abundance, i.e. number of individuals per targeted area~\cite{clark2014more}. 

Most of these algorithms are univariate approaches and focus on individual key species separately~\cite{iverson2019analysis}. However, most species interact with others and their correlation can hold crucial information. This is particularly true for rare species, whose records may be limited in number and borrowing strength from co-occurring species may be fundamental to improve predictions and reduce their uncertainty \cite{clark2014more}. This has motivated the use of, Joint Species Distribution models (JSDM), a group of multivariate approaches that generally use the partial correlation matrix among responses to estimate their interactions \cite{clark2014more}. 

Another key challenge in species distribution modeling is how to address the role of scale on the ecological drivers controlling species distribution~\cite{clark2017generalized}. For example, both SDM and JSDM are generally trained on continental datasets. The resulting models may fail when used to make predictions for local areas, because the same environmental features may have different effects on species abundance locally \cite{gelfand2013scaling}. Understanding and integrating the scaling rules behind changes in features effects will potentially unlock the opportunity to explicitly implement SDMs universally applicable.  Yet, to date we don't even have a complete analysis of the role of scale on joint species distribution modeling.

To address this limit, we studied the relationship of species distributions and species interactions across multiple ecological domains utilizing the generalized joint attribute model (GJAM) \cite{clark2017generalized}. GJAM is a cutting-edge and open source JSDM. Using GJAM for the purpose of this multi-scale analysis has several advantages: (1) it is one of the few joint multilevel model of species abundance available for ecologists, (2) it  integrates partial correlations among species and estimates structure among their interactions \cite{taylor2017joint}; (3) changes in models coefficients can easily be interpreted; (4) it has been used extensively in the last few years, including for predicting tree species abundance at continental scale.

By applying GJAM across different scales we find that species have different interaction patterns and sensitivity to environmental features. We find similar results by investigating different US ecological domains, for species have differing sensitivity to the same climate variables in different zones. 

\section{Methodology}

In this section we will demonstrate a method of modeling tree species  behaviours in different environments. We use GJAM, a JSDM that is well suited for handling multifarious data~\cite{clark2017generalized}. 

Let $Y_i \in \mathbb{R}^S$ represent the abundances of $S$ unique species in the $i^{\text{th}}$ plots, and $X_i \in \mathbb{R}^Q$ be the feature vectors consisting of the $Q$ climate variables corresponding to each plot, for $1 \leq i \leq n$. We model (for the sake of space and simplicity assume all features are continuous):
\begin{equation}
\label{eq:gjam}
    Y_i \sim \, \mathcal{N\!}(B^\top X_i,\, \Sigma)\,
\end{equation}
Where $\mathcal{N(\cdot \raisepunct{,} \cdot)}$ is the multivariate normal distribution, $B \in \mathbb{R}^{Q \times S}$ symbolizes the species response to the climate variables, and $\Sigma \in \mathbb{R}^{S \times S}$ is a covariance matrix that resembles the species interaction between each other. Given observations for $X$ and $Y,$ GJAM estimates $B$ and $\Sigma$ using equation~(\ref{eq:gjam}), we refer the reader to~\cite{clark2017generalized} for further details. With the distribution of species and climate variables the model can predict the species response in a plot.

\subsection{Modeling across Different Ecological Domains}

GJAM has been used to learn a global joint species distribution model~\cite{clark2017generalized} using FIA dataset. We further applied GJAM over different ecological domains defined by the Neon~\cite{neon2017} project by aligning and clustering the FIA data into Neon sites. We use $B$ to study the climate variables' impact on the species across different ecological domains by comparing it from one domain to another. Similarly, \begin{math}\Sigma\end{math} can be used to study species interaction across all domains. Also, having the climate variables the model can predict the species response for plots with no samples or at another point in time.

The environmental predictors we used are slope, aspect (direction a tree is facing due to the slope of the land), elevation, average day length (March, April, October), precipitation (summer months), radiation (August), maximum temperature (August), and minimum temperature (January).

\section{Preliminary Evaluation and Results}
\label{others}

\textbf{Data.} For this study we used the Forest Inventory and Analysis (FIA)\footnote{FIA data available at
\url{https://apps.fs.usda.gov/fia/datamart/datamart.html}.}, an openly available inventory of forests collected at a fine grid nationwide~\cite{bechtold2005enhanced}. Plots are distant from each other ({\tiny \raisebox{0.5ex}{$\sim$}}5km), and cover  an area of ~673 m$^{2}$. Plots are sampled every 5 to 10 years depending on the State.  We used tree species identities from the latest census to have an estimate of the abundances for any target species, at each plot, for current climate. 

The climate data source, Daymet~\cite{Daymet2018v3}, contains daily information at a 1$\times$1 (km) resolution for the climate such as max and min temperature, radiation, and precipitation.

The NEON domains are developed by ecologist, who divided the United States into different ecological regions. The NEON data~\cite{neon2017} has airborne remotely sense images and ground observations of NEON plots.

\subsection{Experimental set up}

The first task is to group the FIA plots into the NEON domains. This process creates 20 separate response and predictor matrices for each of the domains. To address sampling errors driven by small plot size, we group neighbouring plots in clusters of 16 and treat them as contiguous {\tiny \raisebox{0.5ex}{$\sim$}}1ha plots using a same size K-means clustering method based on the coordinates. With the clustered plots, we also aggregated the response matrix so that each observation aligns with the climate observations.

We randomly split 80/20 for train/test sets each of the domains. The evaluation metric for the model’s prediction is the coefficient of determination.

\subsection{Assumptions}

For this experiment, species that do not occur or occur very sparsely in the plots are not predicted; this exclusion may inflate the accuracy of the model. The reason for the exclusion of these species is due to the smaller impact that these species may have on the overall region.

\subsection{Preliminary results}

We have obtain preliminary results for 20 Domains, except Hawaii, Puerto Rico, and Tundra and Taiga in Alaska domains due to the lack of FIA data in those regions. 


\begin{figure}[!tbph]
\centering
  \begin{minipage}[b]{0.31\textwidth}
    \includegraphics[width=\textwidth]{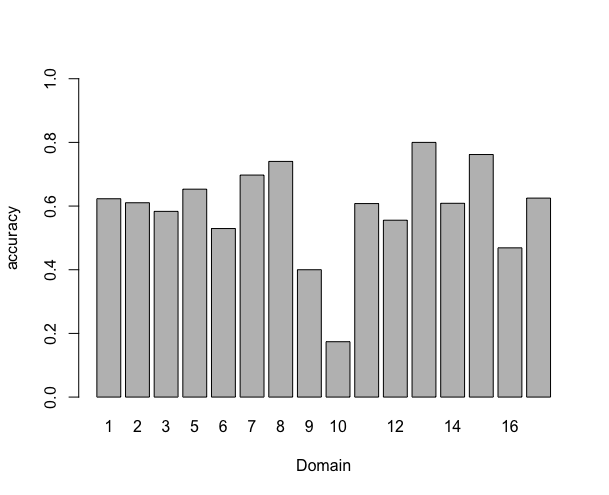}
    \caption{Accuracy per domain, where accuracy is the percent of species where GJAM performed better than the sample mean.}
    \label{figure_1}
  \end{minipage}
  \hspace{3pt}
  \begin{minipage}[b]{0.31\textwidth}
    \includegraphics[width=\textwidth]{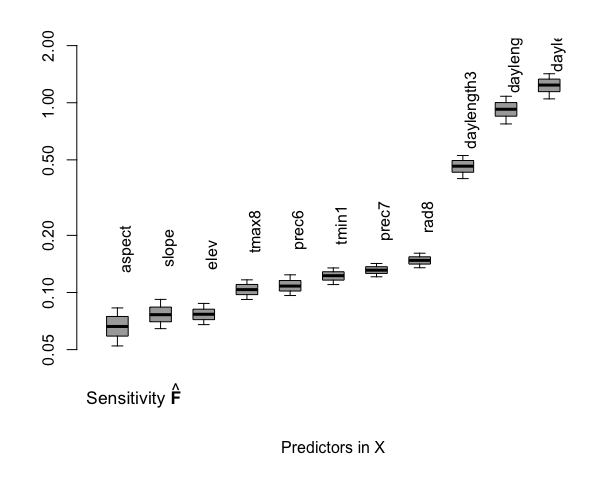}
    \caption{Sensitivity of predictors for Domain 3.}
    \label{figure_2}
  \end{minipage}
  \hspace{3pt}
  \begin{minipage}[b]{0.31\textwidth}
    \includegraphics[width=\textwidth]{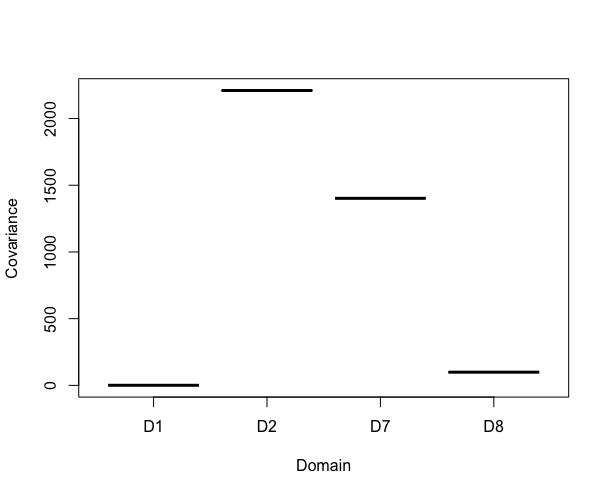}
    \caption{Covariance of Liriodendron tulipifera and Acer Rubrum across Northeast~(D1), Mid Atlantic~(D2), Appalachians~(D7), Ozarks Complex~(D8).}
    \label{figure_3}
  \end{minipage}
\end{figure}

Figure~\ref{figure_2} displays that the day lengths are the most sensitive predictors for domain 3, this is true across all domains. The high importance of \texttt{day length} can be explained through tree phenology, as trees adapt their growing and reproduction cycles over time \cite{chuine2000modelling}. The variation of the impact with other predictors in different domains can be explained by the adaptation to the unique climates, for example elevation has higher impact in mountainous regions such as the Northern Rockies domain compared to the flatter southeast domain.
    
Species performance varies from domain to domain. The reason for this is because a species may occur centrally in one domain while it occurs marginally in another, the model generally performs better when a species occurs centrally.
    
We analyze the covariance of Liriodendron tulipifera and Acer Rubrum which overlap in 4 domains, shown in Figure~\ref{figure_3}. The covariances in domains 2 and 7 are much higher than the other domains. The change in covariances across domains can be seen in other species as well and this variance will be tried to explained through ecological knowledge.

\section{Conclusion and future work}

We outline three publicly available data sources related to climate and tree species. We also apply a straightforward model to predict the abundance of species that takes into account both climate and inter species interactions. This prediction becomes more important with the drastic changes that climate change could have on global ecosystems. Further spatio-temporal analysis of the join species distribution model~\cite{fathony2018distributionally,mollalo2018machine}, to study species distribution shift under climate change over time becomes more important with the drastic changes that climate change could have on global ecosystems. The extension of this work in progress will be enhancing species classification on NEON airborne remote sensing images with information learned from JSDM models. Furthermore we work closely with ecologist to both incorporate rules from domain knowledge in future models and to provide data driven insight back. 

\bibliographystyle{abbrvnat}
\bibliography{neurips_2019_eco}

\end{document}